\documentclass[%
mathleft,%
]{an}
\usepackage{graphicx}
\usepackage{grffile} 
\usepackage{times}
\begin{document}

\Pagespan{1}{}
\Yearpublication{2011}%
\Yearsubmission{2011}%
\Month{1}%
\Volume{999}%
\Issue{92}%

\title{A LOFAR view on the duty cycle of young radio sources}
\author{M. Brienza\inst{1,2}\fnmsep\thanks{Corresponding author:
  \email{brienza@astron.nl}}
\and  R. Morganti\inst{1,2}
\and  A. Shulevski\inst{1}
\and  L. Godfrey\inst{1}
\and N. Vilchez\inst{1}
}
\titlerunning{}
\institute{ASTRON, the Netherlands Institute for Radio Astronomy, Postbus 2, 7990 AA, Dwingeloo, The Netherlands
            \and
             Kapteyn Astronomical Institute, Rijksuniversiteit Groningen, Landleven 12, 9747 AD Groningen, The Netherlands}

\received{July 2015}
\accepted{September 2015}


\abstract{Compact Steep Spectrum, Gigahertz Peaked Spectrum and High Frequency Peak (CSS, GPS, HFP) sources are considered to be young radio sources but the details of their duty cycle are not well understood. In some cases they are thought to develop in large radio galaxies, while in other cases their jets may experience intermittent activity or die prematurely and remain confined within the host galaxy. By studying in a systematic way the presence and the properties of any extended emission surrounding these compact sources we can provide firmer constraints on their evolutionary history and on the timescales of activity of the radio source. Remnant emission from previous outbursts is supposed to have very low surface brightness and to be brighter at low frequency. Taking advantage of the unprecedented sensitivity and resolution provided by the Low Frequency Array (LOFAR) we have started a systematic search of new CSS, GPS and HFP sources with extended emission, as well as a more detailed study of some well-known of these sources. Here we present the key points of our search in the LOFAR fields and a more in-depth analysis on the source B2 0258+35, a CSS source surrounded by a pair of large, diffuse radio lobes. }

\maketitle

\section{Introduction}
\label{intro}

To date, Gigahertz Peak Spectrum, Compact Steep Spectrum and High Frequency Peak (GPS, CSS, HFP) sources are thought to be the best representatives of young radio galaxies because of their compact double morphology which resembles the FRI/FRII extended structures. Both dynamical and radiative arguments suggest that the typical ages of these sources lie in the range $10^2-10^5$ yr (\cite{odea1998}, \cite{owsianik1998}, \cite{murgia1999}). It is expected that in some cases the compact sources will expand to hundreds of kpc sizes on timescales of 1-100 Myr (\cite{parma1999}). However, a significant fraction may never evolve into large sources (\cite{odea1997}, \cite{readhead1996}, \cite{an2012}).

The best way to broaden our knowledge on the duty cycle of these objects is to investigate the presence of extended emission surrounding the compact source. \cite{stanghellini1990} and \cite{stanghellini2005} find respectively that about 25\% (4/15) and 20\% (6/33) of the GPS sources in their samples show associated diffuse emission on the arcsecond-scale, which translates into structures of hundreds of kpc up to few Mpc. The luminosity of this extended emission is often negligible with respect to the compact source. However, its presence can be relevant in terms of the evolutionary history of the radio source.

The most common interpretation of this emission is connected to previous epochs of activity. To date many observations probe that radio galaxies are episodic in nature (\cite{saikia2009}) and statistical studies suggest that this duty cycle strongly depends on the power of the radio source (\cite{best2005}). Sources which exhibit at the same time remnant emitting plasma from previous outbursts and new-born jets are termed \textit{restarted radio galaxies} and can be effectively used to investigate the timescales of activity of the radio AGN. 

One of the most well-known case of GPS with extended emission is the source 0108+388. This source shows a double compact morphology on milli-arcsecond scales as well as some diffuse emission at a distance of 20\arcsec \ from the core connected by a bridge. Multi-frequency studies have been performed by \cite{baum1990} in order to clarify the nature of the source. However, whether the extended plasma is fuelled by the central activity remains unclear. The clearest evidence of restarted activity in CSS and GPS are the sources 3C236 (\cite{willis1974}), B1144+35 (\cite{schoenmakers1999}) and B1245+676 (\cite{marecki2003, saikia2007}). All these sources have Mpc-scale morphologies recalling the structure of a Double-Double Radio Galaxy (DDRG, \cite{schoenmakers1999}) which is suggested to originate from multiple AGN outbursts. 
 
Alternative explanations to the restarting scenario have also been proposed. For example, \cite{baum1990} suggest that GPS/CSS sources could arise within a large-scale radio galaxy if the jet propagation was obstructed or impeded on scales of tens of parsecs to few kiloparsecs. In that case, the extended structure would remain visible along with the confined parsec scale source that would have the appearance of a GPS source. This may occur in coincidence with gas-rich galaxy-galaxy interactions or, more in general, because of instabilities of the accretion process.  

At present, the statistics on young sources with associated extended emission is not sufficient to provide meaningful constraints on their incidence within the general radio galaxy population, and on the implications for the triggering and evolution of radio loud active galaxies. 

Taking advantage of the high sensitivity and high angular resolution at low frequencies provided by the Low Frequency Array (LOFAR, \cite{vanhaarlem2013}) we have started a systematic search of these objects aiming at broaden their statistics. The need for the low frequency is connected to the spectral evolution of the radio emission as the source ages. Indeed, as the injection of fresh particles stops, an exponential break appears in the power-law radio spectrum ($\rm S\propto\nu^{-\alpha}$) at high frequency due to preferential radiative cooling of high-energy electrons (\cite{kardashev1962}, \cite{pacholczyk1970}). Therefore, it is at low frequencies where the source is visible for longer time and where we plan to address our searches. In order to explore the variety of source characteristics and prepare the selection criteria to be used in this search we are also performing single object detailed studies such as the CSS source B2 0258+35, which is surrounded by a pair of large, diffuse radio lobes \cite{shulevski2012}. A combination with higher frequencies will also allow to perform spectral studies to infer the physics of these objects and finally give constraints on their origin and evolution.

In Section 2 we discuss how and why LOFAR promises to be an important instrument in this field, and in Section 3 we present LOFAR observations of B2 0258+35 and discuss possible interpretations for the source. The cosmology adopted in this work assumes a flat universe and the following parameters: $\rm H_{0}= 70\  km \ s^{-1}Mpc^{-1}$, $\Omega_{\Lambda}=0.7, \Omega_{M}=0.3$. 

\section{Searching for AGN remnants with LOFAR}
\label{lofar}

Because of its unprecedented capabilities at low frequency LOFAR (\cite{vanhaarlem2013}) is now playing a primary role in this field of research. The LOFAR Dutch array is currently composed of 24 stations located within a radius of 2 km referred to as the \textit{core} and 14 remote stations arranged in an approximation to a logarithmic spiral distribution extending out to a radius of 90 km. This configuration provides a maximum resolution of $\sim$5\arcsec \ in the high-band (HBA, 110-200 MHz) and $\sim$10\arcsec \ in the low-band (LBA, 30-90 MHz) as well as a good instantaneous uv-coverage on the short baselines. Moreover, the 9 currently operational international stations distributed over 4 different European countries generate baselines up to 1000 km allowing to achieve sub-arcsecond resolutions (e.g. 0.33\arcsec \ at 150 MHz). 

In the context of the search and study of AGN remnant radio plasma there are a number of characteristics that make LOFAR a unique instrument in the field. The first one is its high sensitivity with noise values of 0.2 mJy at 150 MHz and 3.9 mJy at 60 MHz for 10 hours observations. This is essential for detecting low-surface brightness sources like AGN remnant radio plasma. The second one is the variety of resolutions that are provided by a single observation. This unique array configuration, which combines a dense core of short baselines together with longer baselines, allows at the same time a good imaging of diffuse emission as well as compact, small-scale structures providing a complete morphological characterization of the sources. Moreover, LOFAR highest resolutions match for the first time those at higher frequency and open the way to systematic resolved spectral studies. Finally, the wide field of view produced by LOFAR observations (e.g. 6 deg at 150 MHz and 12 deg at 60 MHz) allows significant statistical studies and source searches directly in the surroundings of the target source.  

A few detections of remnant and restarted radio galaxy in the LOFAR fields have already been presented in literature (NGC 5590, \cite{vanweeren2014}; 4C35.06, \cite{shulevski2015}; J1828+49, \cite{brienza2015}) providing a taste of the instrument capabilities. These sources have been morphologically identified based on their amorphous shapes and low-surface brightness of a few mJy $\rm arcmin^{-2}$. This class of objects is expected to be numerous in the LOFAR fields and therefore new systematic searches are being planned. 

One of the Key Science Project of LOFAR is the continuum surveys which follow a three-tier approach with different observational setups (\cite{rottgering2003}, \cite{vanhaarlem2013}). The Tier-1 high-frequency survey will cover the entire northern sky with a resolution of 5\arcsec \ and an rms of 0.07 mJy at frequencies 120-180 MHz. As to June 2015 the high-frequency Tier-1 survey has completed about 1000 $\rm deg^2$ observations.  Typical maps have 20\arcsec \ resolution and noise equal to 0.5 mJy $\rm beam^{-1}$ with the only limitation to the image quality being the ionospheric calibration. A low-frequency Tier-1 survey is also scheduled to start in the next future.

Using the fields already observed and reduced for the surveys we have started a search of remnant and restarted radio galaxies that will be further expanded to the whole northern sky in the future. Beyond blind searches we aim to systematically investigate the surroundings of all the well-known CSS/GPS/HFP sources in order to provide a better statistic of the presence of associated extended emission and investigate its nature in closer detail. An initial search in this direction has already been performed by \cite{shulevski2015thesis} using the LOFAR Multifrequency Snapshot Sky Survey (MSSS, \cite{heald2015}). 
However, the sensitivity and spatial resolution of this survey turned out to be not ideally suited for this search and no new detections were made even though spectral index studies suggested that some CSS sources might show an excess of emission. Despite that, MSSS nicely detects two of the well-known giant radio galaxies with GPS or CSS central cores mentioned in Sec. \ref{intro} (i.e. 3C236 and B1245+676). These issues are overcome by the high resolution and sensitivity provided by the LOFAR Tier-1 survey which can probe extended emission on 5-10 arcsecond scales. Further details about the selection criteria we intend to use for identifying remnant radio plasma are discussed in \cite{brienza2015}. 
Higher frequency data will be then required to complete the radio spectral coverage allowing radiative ageing analysis of the sources. Furthermore, complementary data in other frequency regimes will probe the properties of the host galaxy and its environment.

In this way we aim at creating new representative samples of remnant and restarted radio galaxies to start a systematic investigation of their properties as a function of energetics, host galaxy and environment.

\section{B2 0258+35: a restarting radio galaxy?}

In the context of restarting radio galaxies the CSS source B2 0258+35 has revealed a very intriguing story. The radio source has a power equal to $\rm 10^{24.37} \ W\ Hz^{-1}$ at 408 MHz (\cite{giovannini2001}) and it is associated to the field early-type galaxy NGC 1167 (z=0.0165) which is optically classified as a Seyfert 2 galaxy (\cite{ho1997}). Its radio morphology was first studied by \cite{sanghera1995} using the VLA and MERLIN+EVN at 5 and 1.6 GHz respectively. They classified it as a CSS source even though the high frequency spectrum is not very steep $\alpha = 0.54$. 

\begin{figure}[h!]
\centering
\includegraphics[width=0.5\textwidth]{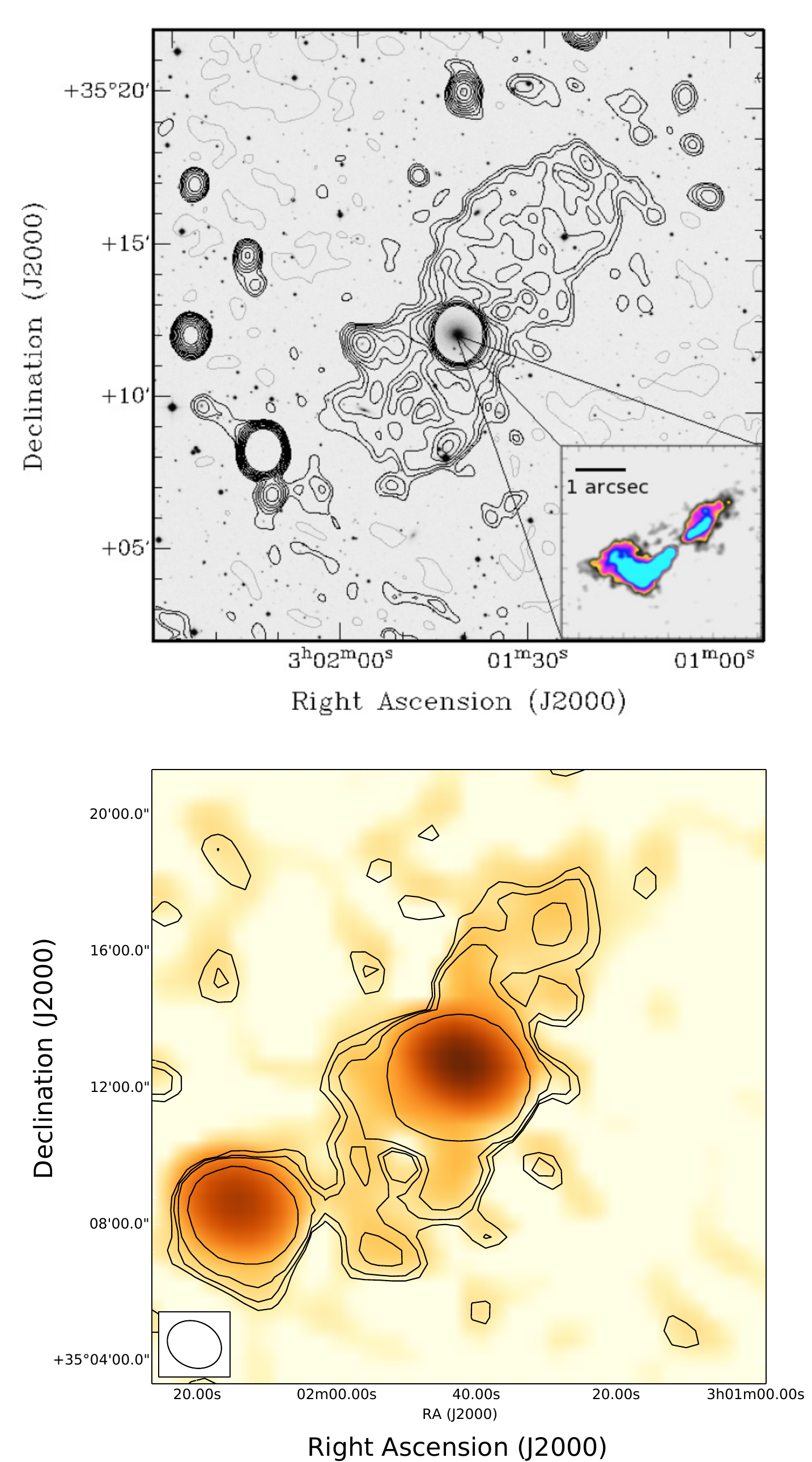}
	
\caption{Radio maps of the source B2 0258+35. \textit{Top -} 1.4-GHz map from \cite{shulevski2012} with zoom-in on the CSS source from \cite{giroletti2005}. \textit{Bottom -} LOFAR 145-MHz map. Levels: 2, 3, 5, 20 $\rm \times \sigma (= 3 \ mJy \ beam^{-1}$)}.
\label{fig:map}
\end{figure}

\cite{giroletti2005} provide better imaging of the source using the VLA at 8 and 22 GHz. At sub-arcsecond resolution the source shows a two-lobe morphology of 3.2\arcsec \ size (1.1 kpc) without any clear indication of hot-spots and the southern lobe exhibits an extension towards north-east (see inset in Fig. \ref{fig:map} in the top panel). At milli-arcsecond resolution \cite{giovannini2001} further detect a compact component with jet-like structure. Assuming an equipartition magnetic field of B = $9\times 10^{-5}$ Gauss and the detected break frequency at 4.6 GHz, \cite{giroletti2005} compute an average spectral age of $9\times 10^5$ yr. It is worth noting though that the outermost regions of the lobes show a quite steep spectral index in the range 1.0-1.5. \cite{giroletti2005} speculate that the source will not grow into an FRI/II radio galaxy because i) it doesn't show hot-spots, ii) the bending of the southern lobe indicates a dense surrounding ISM, iii) its size is small for the estimated spectral ages.

A new significant ingredient for reconstructing the evolutionary history of the source is the discovery by \cite{shulevski2012} of an extended emission with double-lobe morphology of 240 kpc size surrounding the CSS (see Fig.\ref{fig:map} top panel). The very low surface brightness emission at 1.4 GHz with average value of 1.4 mJy $\rm arcmin^{-2}$ remained undetected in all of the previous studies. The low surface brightness and lack of compact features such as hotspots or jets likely indicate that the emission could be remnant plasma from a previous cycle of activity (\cite{shulevski2012}). Assuming the lobes to be buoyant bubbles expanding in the intergalactic medium \cite{shulevski2012} estimate a time since the last acceleration of particles in the outer lobes of about 100 Myr. Because the age of the young central CSS source is about $\rm 9 \times 10^5$ yr they suggest that the quiescent phase between the two radio bursts lasted about 100 Myr.

\subsection{LOFAR observations of B2 0258+35}

In order to further investigate the nature of the source we performed LOFAR observations in the HBA. The source was observed for 8 hours using the Dutch array with a bandwidth of 40 MHz. 3C48 has been used as flux density calibrator and the flux scale has been set according to Scaife \& Heald (2012). The amplitude corrected visibilities have been phase self-calibrated and the final image was produced using a robust weighting of -0.3 and has final beam size of 98 $\times$ 80 arcsec. The RMS of the map is 3 mJy beam$^{-1}$. 

The integrated spectrum of the central CSS source is presented in Fig. 2 with the LOFAR measurements at 120 MHz and 160 MHz nicely following the literature points. The linear relation between size and turnover frequency found by \cite{odea1997} predicts a turnover in the CSS spectrum at an observed frequency of 570 MHz. 
The observed spectral shape suggests that the turnover frequency is at $\nu \leq 300$ MHz. It is also worth noting that at lower frequency there may be contamination to the CSS flux density from the surrounding extended emission. Future high resolution LOFAR LBA observations will trace the spectrum at low frequency in more detail giving constraints on the spectral peak. 

\begin{figure}[h!]
\centering
\includegraphics[width=0.53\textwidth]{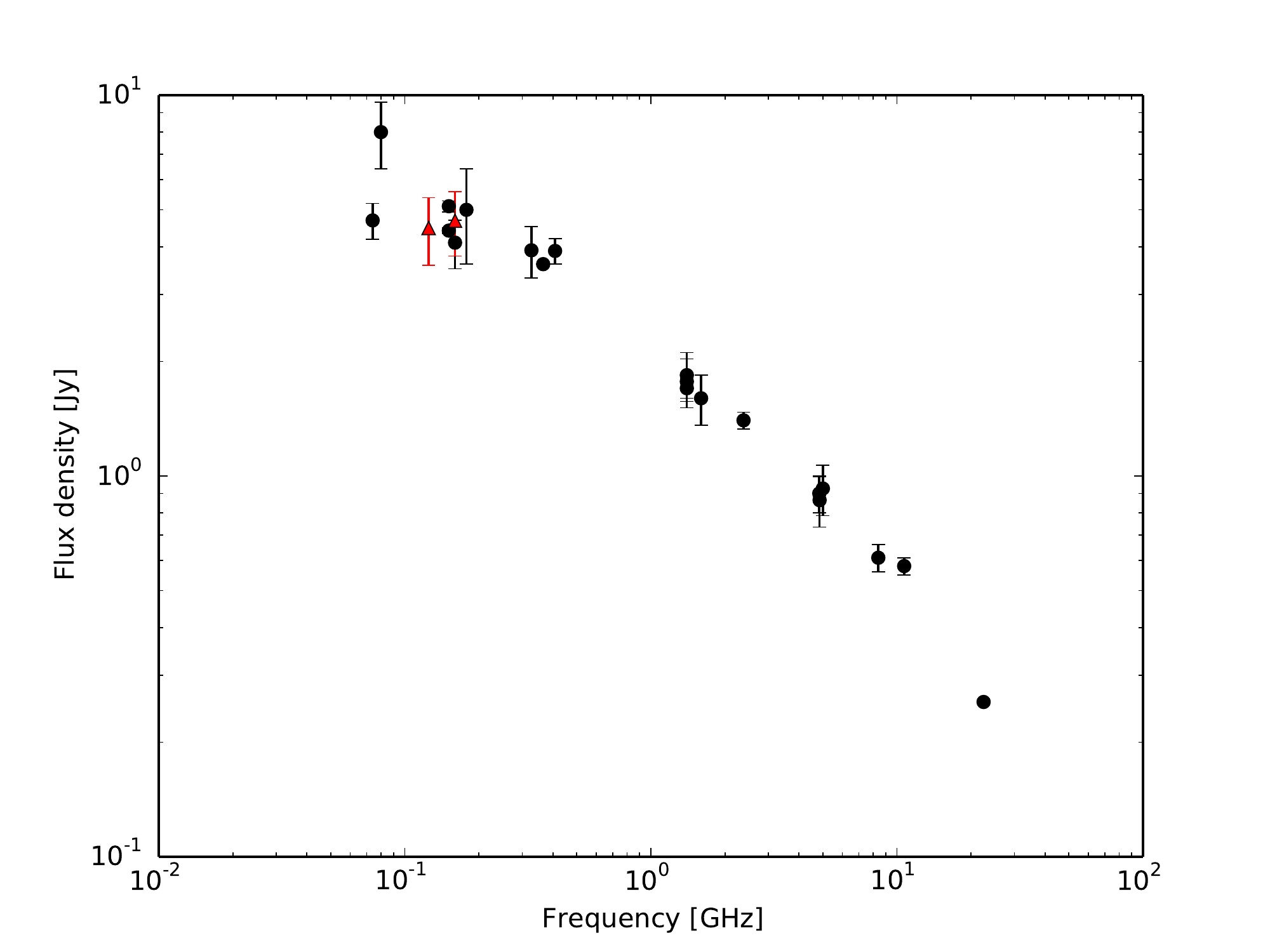}
\caption{The integrated radio spectrum of the B2 0258+35 CSS source is shown. Red triangles represent the new LOFAR flux density measurements at 120 MHz and 160 MHz while blue circles are taken from literature (\cite{giroletti2005}, NVSS, WENSS, NED)}
\label{fig:spectrum}
\end{figure}

The morphology of the extended emission as seen by LOFAR at 145 MHz reflects what was previously observed at 1.4 GHz by \cite{shulevski2012} (see Fig. \ref{fig:map}). The linear extension of the lobes at low frequency is in agreement with the high frequencies and the surface brightness has an average value of 4.7 mJy arcmin$^{-2}$. The two broad enhancements in surface brightness observed in the Northern and Southern direction are clearly visible at both high and low frequencies but we cannot interpret univocally their origin.

The extended emission that we see has likely been originated by large-scale jets that are currently switched off or smothered. As discussed in Sec. \ref{lofar} the spectrum of an ageing plasma is expected to steepen because of radiative cooling to typical spectral indices $\alpha\geq1.2$. Surprisingly, a preliminary spectral index study of the extended lobes of B2 0258+35 reveal a non-steep spectrum in the range 145-1400 MHz with values $\rm \alpha_{1400}^{145}=0.6-0.8$. 

Two scenarios can be invoked in order to explain our findings of a relatively "normal" spectral index for the extended lobes, when a much steeper spectrum typical of aged plasma may have been expected.
On one hand the outer lobes of B2 0258+35 could be completely detached from the current central activity but they could manifest their spectral steepening only at frequencies higher than 1.4 GHz, therefore beyond our available spectral window. This occurrence has already been verified in a few remnant radio galaxies e.g. B2 1610+29 (\cite{murgia2011}) and J1828+49 (\cite{brienza2015}) and can be explained by a combination of remnant plasma energetics and possibly local environment. These sources show low magnetic field values equal to 3.2 $\rm \mu G$ and 1 $\rm \mu G$ respectively and do not suffer significant expansion energy losses so that the spectral shape at low frequency is still comparable to those observed in active sources. In this way, they remain detectable with spectral steepening occurring only at high frequency, for a long time after the jets switch off. 

On the other hand, the outer lobes of B2 0258+35 may still be fuelled by the current AGN activity and consequently the spectrum has not developed a high energy cutoff. Because we do not recognize any defined large-scale jets this could be happening via decollimated jets and likely at a low rate. The best example of this phenomenon is observed Centaurus A. This radio galaxy with clear signs of intermittent jet activity (\cite{morganti1999}) likely due to a merger event, shows a relevant difference between the northern and southern outer lobe. The first one indeed exhibits an intermediate-scale northern middle lobe (\cite{morganti1999}) which is interpreted as an ``open channel" between the inner and the outer lobe through which fresh injected particles can reach the outskirts. This idea is supported by the fact the the entire northern lobe does not present any signature of spectral steepening up to 90 GHz (\cite{hardcastle2009}). Although this last scenario might explain what we observe in B2 0258+35, the reason why the jets would have been smothered remains unclear. \cite{struve2010} show indeed that the last merger the host galaxy has experienced goes back up to 1 Gyr, hence incompatible with the radio activity timescales.  

\section{Conclusions and future perspective}

B2 0258+35 is a rare example of a CSS source with large low surface brightness radio lobes surrounding it. The extended lobes may be the remnants from a previous cycle of jet activity that arise due to intermittency of the central engine. LOFAR nicely detects this low surface brightness structure at 145 MHz and allows a first investigation of its spectral properties in combination with 1.4 GHz WSRT data. Contrary to expectations, we find that the extended lobes do not show very steep spectral index as predicted for old ageing plasma.
If the lobes are actually remnants of a previous cycle of activity, the spectral steepening must occur at frequencies higher than 1.4 GHz. Otherwise they could still being fueled from the central engine via an open channel between the inner and outer lobes through which freshly injected particles can reach the outskirts. Upcoming 320 MHz and 5 GHz observations will provide new constraints on the radio spectrum of the extended emission enabling a more thorough spectral ageing study, from which firmer conclusions may be drawn. 

This study confirms the idea that large scale diffuse emission surrounding compact sources can show a variety of physical properties and may have experienced a variety of evolutionary paths. A clearer understanding of the spectral properties of the remnant plasma surrounding CSS, GPS and HFP sources is crucial for the identification and selection of this class of objects from the LOFAR survey. For this reason it is important to investigate in more details well-known sources in preparation for all-sky selections. Upcoming complete samples of CSS, GPS and HFP with extended emission will finally allow a more systematic study of the young radio sources duty cycle.

\bigskip

\acknowledgements
We thank the organizers of the 5th workshop on CSS and GPS radio sources and RadioNet for the financial support. 
The research leading to these results has received funding from the European Research Council under the European Union's Seventh Framework Programme (FP/2007-2013) / ERC Advanced Grant RADIOLIFE-320745. LOFAR, the Low Frequency Array designed and constructed by ASTRON (Netherlands Institute for Radio Astronomy), has facilities in several countries, that are owned by various parties (each with their own funding sources), and that are collectively operated by the International LOFAR Telescope (ILT) foundation under a joint scientific policy. This research has made use of the NASA/IPAC Extragalactic Database (NED), which is operated by the Jet Propulsion Laboratory, California Institute of Technology,
under contract with the National Aeronautics and Space Administration. This research made use of APLpy, an open-source plotting package for Python hosted at http://aplpy.github.com.


\begin{thebibliography}{}
\bibitem[An \& Baan (2012)]{an2012} An, T., \& Baan, W.~A.\ 2012, ApJ, 760, 77 
\bibitem[Baum et al. (1990)]{baum1990} Baum, S.~A., O'Dea, C.~P., Murphy, D.~W., \& de Bruyn, A.~G.\ 1990, A\&A, 232, 19 
\bibitem[Best et al. (2005)]{best2005} Best, P.~N., Kauffmann, G., Heckman, T.~M., et al.\ 2005, MNRAS, 362, 25 
\bibitem[Brienza et al. (2015)]{brienza2015} Brienza, M., Godfrey, L.,Morganti, R., et al. 2015 accepted by A\&A 
\bibitem[Owsianik \& Conway (1998)]{owsianik1998} Owsianik, I., \& Conway, J.~E.\ 1998, ApJ, 337, 69 
\bibitem[Giovannini et al. (2001)]{giovannini2001} Giovannini, G., Cotton, W.~D., Feretti, L., Lara, L., \& Venturi, T.\ 2001, ApJ, 552, 508 
\bibitem[Giroletti et al. (2005)]{giroletti2005} Giroletti, M., Giovannini, G., \& Taylor, G.~B.\ 2005, A\&A, 441, 89 
\bibitem[Hardcastle et al. (2009)]{hardcastle2009} Hardcastle, M.~J., Cheung, C.~C., Feain, I.~J., \& Stawarz, {\L}.\ 2009, MNRAS, 393, 1041  
\bibitem[Heald et al. (2015)]{heald2015} Heald, G.~H., Pizzo, R.~F., Orr{\'u}, E., et al.\ 2015, A\&A, 582, A123
\bibitem[Ho et al. (1997)]{ho1997} Ho, L.~C., Filippenko, A.~V., \& Sargent, W.~L.~W.\ 1997, ApJ, 112, 315 
\bibitem[Kardashev (1962)]{kardashev1962} Kardashev, N.~S.\ 1962, Soviet Ast., 6, 317 
\bibitem[Marecki et al. (2003)]{marecki2003} Marecki, A., Barthel, P.~D., Polatidis, A., \& Owsianik, I.\ 2003, PASA, 20, 16 
\bibitem[Morganti et al. (1999)]{morganti1999} Morganti, R., Killeen, N.~E.~B., Ekers, R.~D., \& Oosterloo, T.~A.\ 1999, MNRAS, 307, 750 
\bibitem[Murgia et al. (1999)]{murgia1999} Murgia, M., Fanti, C., Fanti, R., et al.\ 1999, ApJ, 345, 769 
\bibitem[Murgia et al. (2011)]{murgia2011} Murgia, M., Parma, P., Mack, K.-H., et al.\ 2011, A\&A, 526, A148 
\bibitem[O'Dea \& Baum (1997)]{odea1997} O'Dea, C.~P., \& Baum, S.~A.\ 1997, AJ, 113, 148 
\bibitem[O'Dea (1998)]{odea1998} O'Dea, C.~P.\ 1998, PASP, 110, 493  
\bibitem[Pacholczyk (1970)]{pacholczyk1970} Pacholczyk, A.~G.\ 1970, Series of Books in Astronomy and Astrophysics, San Francisco: Freeman, 1970
\bibitem[Parma et al. (1999)]{parma1999} Parma, P., Murgia, M., Morganti, R., et al.\ 1999, A\&A, 344, 7 
\bibitem[Readhead et al. (1996)]{readhead1996} Readhead, A.~C.~S., Taylor, G.~B., Pearson, T.~J., \& Wilkinson, P.~N.\ 1996, ApJ, 460, 634 
\bibitem[R{\"o}ttgering et al. (2003)]{rottgering2003} R{\"o}ttgering, H., de Bruyn, A.~G., Fender, R.~P., et al.\ 2003, Texas in Tuscany.~XXI Symposium on Relativistic Astrophysics, 69 
\bibitem[Saikia et al. (2007)]{saikia2007} Saikia, D.~J., Gupta, N., \& Konar, C.\ 2007, MNRAS, 375, L31 
\bibitem[Saikia 
\& Jamrozy (2009)]{saikia2009} Saikia, D.~J., \& Jamrozy, M.\ 2009, Bulletin of the Astronomical Society of India, 37, 63 
\bibitem[Sanghera et al. (1995)]{sanghera1995} Sanghera, H.~S., Saikia, D.~J., Luedke, E., et al.\ 1995, A\&A, 295, 629 
\bibitem[Schoenmakers et al. (1999)]{schoenmakers1999} Schoenmakers, A.~P., de Bruyn, A.~G., R{\"o}ttgering, H.~J.~A., \& van der Laan, H.\ 1999, A\&A, 341, 44 
\bibitem[Shulevski et al. (2012)]{shulevski2012} Shulevski, A., Morganti, R., Oosterloo, T., \& Struve, C.\ 2012, A\&A, 545, A91 
\bibitem[Shulevski (PhD thesis, 2015)]{shulevski2015thesis} Shulevski, A. 2015, PhD thesis, RuG
\bibitem[Shulevski et al. (2015)]{shulevski2015} Shulevski, A., Morganti, R., Barthel, P.~D., et al.\ 2015, arXiv:1504.06642 
\bibitem[Stanghellini et al. (1990)]{stanghellini1990} Stanghellini, C., Baum, S.~A., O'Dea, C.~P., \& Morris, G.~B.\ 1990, A\&A, 233, 379 
\bibitem[Stanghellini et al. (2005)]{stanghellini2005} Stanghellini, C., O'Dea, C.~P., Dallacasa, D., et al.\ 2005, A\&A, 443, 891 
\bibitem[Struve et al. (2010)]{struve2010} Struve, C., Oosterloo, T., Sancisi, R., Morganti, R., \& Emonts, B.~H.~C.\ 2010, A\&A, 523, A75 
\bibitem[van Haarlem et al. (2013)]{vanhaarlem2013} van Haarlem, M.~P., Wise, M.~W., Gunst, A.~W., et al.\ 2013, A\&A, 556, A2 
\bibitem[van Weeren et al. (2014)]{vanweeren2014} van Weeren, R.~J., Williams, W.~L., Tasse, C., et al.\ 2014, ApJ, 793, 82 
\bibitem[Willis et al. (1974)]{willis1974} Willis, A.~G., Strom, R.~G., \& Wilson, A.~S.\ 1974, Nat, 250, 625 


\end{thebibliography}
\end{document}